\documentclass[10pt, conference, letterpaper]{IEEEtran}

\usepackage{framed}
\usepackage{hyperref}
\usepackage{tcolorbox}
\usepackage{cite}
\usepackage{amsmath}
\usepackage{amsfonts}
\usepackage{mathtools}
\usepackage{amssymb}
\usepackage{amsthm}
\usepackage[ruled,linesnumbered,noend]{algorithm2e}
\usepackage{cases}
\usepackage{graphicx}
\usepackage{bbm}
\usepackage{subcaption}
\usepackage{tabularx}

\IEEEoverridecommandlockouts

\begin{document}
%
\title{Achieving AoI Fairness in Spatially Distributed Wireless Networks: From Theory to Implementation}
\author{Nicholas Jones, Joshua Wornell, Chao Li, and Eytan Modiano\\
\thanks{The authors are with the Laboratory for Information and Decision Systems (LIDS), Massachusetts Institute of Technology, Cambridge, MA 02139.\newline email: \{jonesn, jwornell, chaoli, modiano\}@mit.edu}}
\vspace{-3mm}

\IEEEaftertitletext{\vspace{-0.6\baselineskip}}
\maketitle

\begin{abstract}
    We design and implement two variants of a practical random access protocol called WiFair, based on IEEE 802.11 and designed to mitigate spatial unfairness in Age of Information (AoI). Drawing on previous theoretical work, we modify the mechanics of 802.11 to fairly minimize AoI in a wireless network consisting of several update nodes and a single base station. We implement this protocol on a testbed of software defined radios (SDRs) and measure its performance under a variety of settings compared to standard 802.11. We observe a $32\%$ reduction in network average AoI and an $89\%$ reduction in peak AoI in a last come first served (LCFS) single-packet queue setting, as well as a $76\%$ reduction in network average AoI and an $82\%$ reduction in peak AoI in a first come first served (FCFS) queue setting when the network is congested. We further show that when the network is uncongested, WiFair achieves the same performance as 802.11, and we demonstrate its robustness to more bursty traffic by streaming live video.
\end{abstract}

\maketitle

\section{Introduction}~\label{sec:intro}

Future wireless networks will require support for time-sensitive monitoring and control. Robots in automated warehouses and autonomous vehicles on the road will rely on fresh information from neighboring devices to make decisions that optimize performance and ensure the safety of those around them. A metric known as Age of Information (AoI) has been used to quantify the information freshness of each neighboring device from the perspective of the receiver~\cite{kaul2012real}. AoI measures the time that has elapsed since the last update received from each node was generated. When the AoI of each node is small, the receiver has fresh information from the entire network and can make appropriate decisions.

There has been considerable work done on scheduling to minimize AoI in wireless networks. The most mature setting assumes a centralized controller that chooses a scheduling policy to minimize some function of AoI. In~\cite{sun2017update}, the authors show that in a single-source network, a zero-wait policy is not always optimal. In~\cite{bedewy2019minimizing}, the authors examine AoI in the presence of queueing. More settings are considered by optimizing AoI under general interference constraints in~\cite{talak2020optimizing}, for broadcast networks in~\cite{kadota2018scheduling}, and for networks with throughput constraints in~\cite{kadota2019scheduling}. Finally,~\cite{tripathi2019whittle} considers general functions of AoI.

There has also been work done on optimizing random-access schemes for AoI in the absence of a centralized scheduler. In~\cite{yates2017status}, the authors optimize slotted ALOHA to minimize weighted sum AoI, and in~\cite{kadota2021age}, the authors optimize both slotted ALOHA and CSMA in the presence of stochastic arrivals. In~\cite{maatouk2020age}, CSMA is optimized when updates are generated at will, and~\cite{tripathi2023fresh} shows that by incorporating AoI into the CSMA backoff timers, the network will mimic a near-optimal centralized policy with high probability. In~\cite{chen2022age} and~\cite{ahmetoglu2022mista}, AoI-threshold policies are proposed, where each node only participates in a random access scheme when its age exceeds some value.

While most of the work on AoI is theoretical in nature, a few works have implemented protocols and tested real-time performance using software defined radios (SDRs). In~\cite{kadota2021wifresh}, the authors implement both a MAC layer and application layer protocol for minimizing AoI, based on IEEE 802.11 (standard WiFi). In~\cite{tripathi2023wiswarm}, an application layer protocol is implemented to minimize AoI, and results are shown using a live control loop with UAVs performing object tracking. In~\cite{han2020software}, the authors implement a full-scale SDR prototype and test several theoretical AoI-minimizing policies. An AoI-aware downlink scheduling policy is implemented on an SDR testbed in~\cite{oguz2022implementation}. In~\cite{shreedar2022coexistence}, the authors study how prioritization at the MAC layer affects AoI in the presence of both age-sensitive and high throughput traffic. In~\cite{shreedar2024acp}, a transport layer protocol is implemented to minimize AoI by regulating traffic rates. Finally, the authors of~\cite{ayan2021experimental} implement a centralized scheduling policy on SDRs to measure AoI under various queueing methods, and obtain results using a real control system with inverted pendulums.

In each of the theoretical works on AoI above, the authors assume a collision model, where an update is received by the base station only if no other updates are sent at the same time. In each of the implementations above, the protocols are designed using the same assumption. For networks with~\textit{spatial diversity}, where some nodes are much closer to the base station or transmit with higher power than others, the collision model is an oversimplification. In reality, the stronger signal will be received, and only the weaker signal will experience a collision.

A more realistic model for this setting is the capture model, where an update is received if its signal to interference (SIR) ratio exceeds a known threshold. In our recent work~\cite{jones2023minimizing}, we consider a random access scheme which uses the capture model and takes into account the spatial configuration of the network. Following a similar line of reasoning to~\cite{celik2009mac}, we show that under traditional random access schemes where nodes are treated identically, the network can experience significant \textit{spatial unfairness} in terms of AoI. To address this issue, we propose policies where each node transmits with a unique probability, which is optimized for minimizing AoI under different fairness metrics, and takes into account the network's spatial configuration. We show that when nodes which are farther from the base station transmit more frequently, their individual AoI and the average AoI of the network is significantly improved.

In this work, we design and implement two practical and spatially fair random-access protocols on a testbed of SDRs. The protocols are based on IEEE 802.11 (standard WiFi), but we modify the transmission mechanics to emulate our policies in~\cite{jones2023minimizing}. In this way, we leverage the practical aspects of 802.11, while mitigating its spatial unfairness and achieving near-optimal AoI. We measure the real over-the-air performance of our protocols against standard 802.11 in three separate experiments:
\begin{itemize}
    \item A last come first served (LCFS) single-packet queue setting, which is the model used in~\cite{jones2023minimizing}.
    \item A first come first served (FCFS) queue setting where the sources send live UDP video streams.
    \item A FCFS queue setting where the network is congested.
\end{itemize}

We show that in the LCFS single-packet queue case, and the FCFS queue case when the network is congested, our protocols drastically improve AoI compared to standard 802.11. In the second experiment, we introduce less congestion to preserve the quality of the UDP streams. We observe similar results to 802.11, while demonstrating that our protocols can serve burstier traffic like a video stream without loss of performance.



\section{Theoretical Background}\label{sec:theoretical-policies}

We begin by introducing the theoretical background from~\cite{jones2023minimizing} and formally defining Age of Information. We consider a network of $N$ sources sending status updates wirelessly to a central base station. Time is slotted, and in each slot one or more sources can attempt to send an update. The duration of a time slot is equal to the time it takes to send one update packet over the wireless channel. All sources share the same channel, and so are subject to interference. We assume for simplicity of analysis that noise is negligible relative to interference, and so is ignored.

Let $\mu_i(t) = 1$ if the base station receives an update from node $i$ at time $t$, and $\mu_i(t) = 0$ otherwise. Denote $\tau_i$ as the time when the last update received at the base station from node $i$ was generated. Let each source have a LCFS single-packet queue, and assume that each update consists of a single packet, and that each node samples an update immediately before transmitting. This ensures that each update received by the base station contains the most fresh information, and it follows that $\tau_i$ is simply the time the last update was received from node $i$.

\begin{figure}
    \centering
    \includegraphics[width=0.5\textwidth]{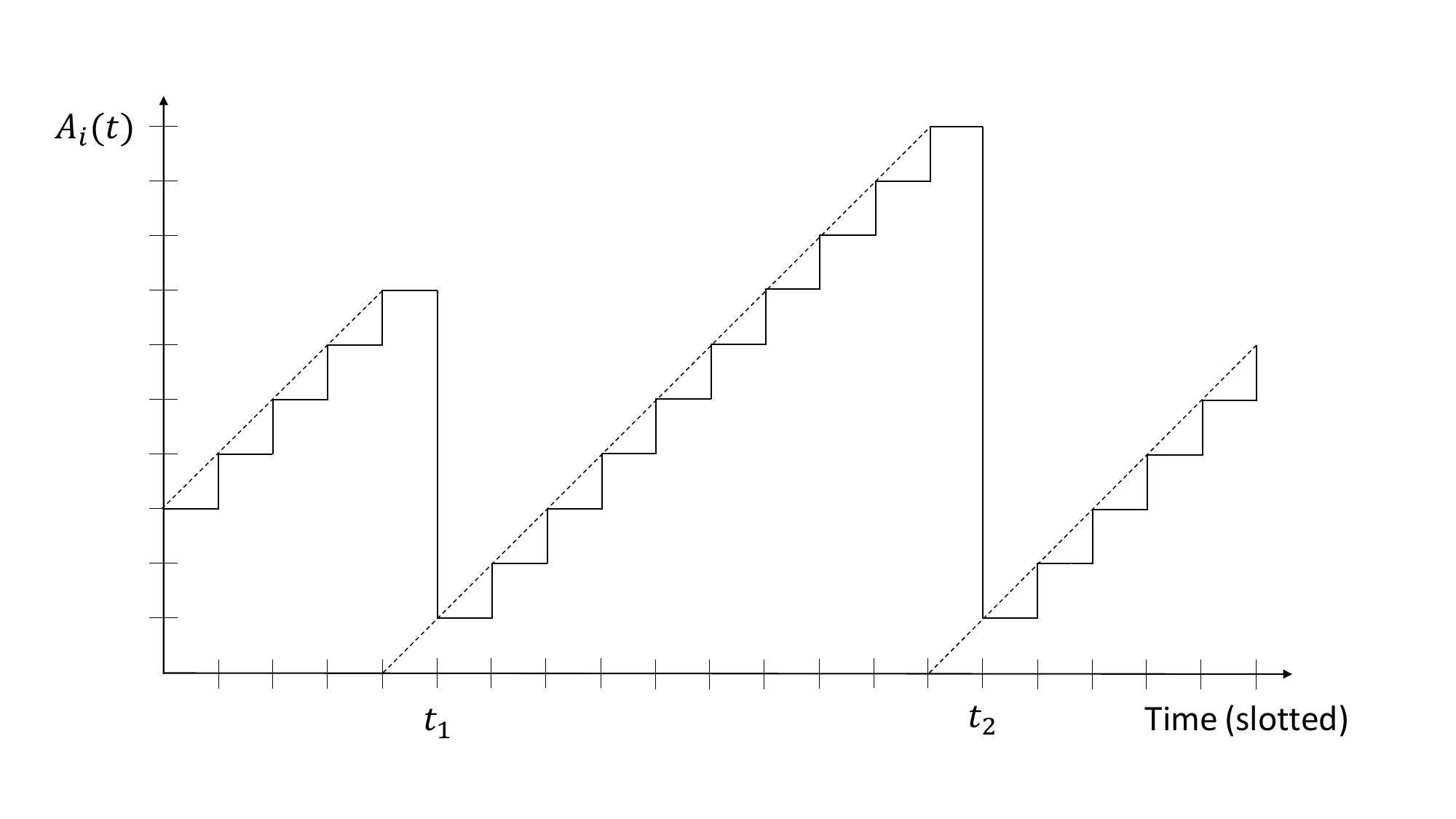}
    \caption{AoI evolution over time}
    \label{fig:aoievolution}
\end{figure}

Define the AoI of node $i$ at time $t$ as 
\begin{equation}
    A_i(t) \triangleq t - \tau_i,
\end{equation}
and let it denote the time that has elapsed since the last packet received at the base station from node $i$ was generated. Note that it evolves as 
\begin{equation}
    A_i(t+1) = \begin{cases}
        A_i(t) + 1, &\mu_i(t) = 0, \\
        1, &\mu_i(t) = 1,
    \end{cases}
\end{equation}
and so follows a sawtooth pattern. This is shown graphically in Figure~\ref{fig:aoievolution}, where $t_1$ and $t_2$ each represent time slots where $\mu_i(t) = 1$. Finally, define the time average AoI of node $i$ as 
\begin{equation}
    h_i \triangleq \lim_{T \to \infty} \frac{1}{T} \sum_{t=1}^T \mathbb{E}[A_i(t)].
\end{equation}
This limit is shown to exist in~\cite{jones2023minimizing}.

In every time slot, each source $i$ transmits with some static probability $p_i$, independently of other sources and across time. The goal of the theoretical policies in~\cite{jones2023minimizing} is to optimize these probabilities to minimize some function of AoI. We assume a Rayleigh fading channel and a fixed, known rolloff paramter $\beta$, such that the signal strength of source $i$ after fading decreases as $r_i^{-\beta}$, where $r_i$ is the distance of node $i$ from the base station, normalized to take values from $0$ to $1$. Finally, we assume that when a source sends an update, it is received at the base station if its SIR exceeds a known, fixed threshold $\theta$, which is a function of the modulation and coding (MCS) scheme.

Under this model, the time average AoI is shown to be 
\begin{equation}
    h_i = \frac{1}{p_i \prod_{j \in I_i} \big(1 - \frac{p_j}{1 + d_{ij}} \big)},
\end{equation}
where $I_i$ is the set of nodes which interfere with node $i$, and $d_{ij} = r_i^{-\beta}/(r_j^{-\beta} \theta)$. Note that $r_i^{-\beta}$ is the average received power at the base station normalized by the transmission power.

In this work, we focus on proportionally fair AoI, which is a direct analog of proportionally fair throughput~\cite{kelly1997charging}, and is the solution to 
\begin{align}\label{eq:pfopt}
\begin{aligned}
    \min_{\boldsymbol{p}} \ &\sum_{i=1}^N \log h_i, \\
        \text{s.t.} \ &0 \leq p_i \leq 1, \ \forall \ i.
\end{aligned}
\end{align}
We focus on this metric as opposed to min sum AoI or min max AoI because the optimal policy is completely separable, meaning each source can compute its optimal transmission probability independently. This naturally lends itself to a distributed implementation, and we show in~\cite{jones2023minimizing} that the performance is very similar to min sum AoI.

We consider two separate cases of proportionally fair AoI. In the first case, denoted as the proportionally fair (PF) policy, we assume that each source has full knowledge of the network topology. In the second case, denoted as the topology agnostic (TA) policy, we assume that each source is only aware of its own location, and that sources are distributed uniformly in a two-dimensional disc around the base station.

The PF policy is the solution to~\eqref{eq:pfopt}, and is given by $p_i^{PF} = \min \{ \tilde{p_i}^{PF}, 1 \}$ for all $i$, where $\tilde{p_i}^{PF}$ is the solution to the fixed point equation
\begin{equation}\label{eq:pfsolution}
    \frac{1}{\tilde{p_i}^{PF}} - \sum_{j \in I_i} \frac{1}{1+d_{ji} - \tilde{p_i}^{PF}} = 0.
\end{equation}
Note that each source $i$ can comupte $p_i^{PF}$ independently given knowledge of $d_{ji}$ for each interferer $j$. The fixed point equation~\eqref{eq:pfsolution} converges, and thus can be solved iteratively. Finally, note that $\tilde{p_i}^{PF} \geq 1$ only when $d_{ji}$ is large for each interferer $j$, which signifies that source $i$ has a much smaller average received power. In general, the weaker source $i$'s signal is relative to its interferers, the larger $p_i^{PF}$ becomes. This illustrates the idea that when some sources' transmissions dominate others, the weaker sources can be much more aggressive, reducing the number of idle slots where no one transmits without impacting the stronger transmissions. 

Similarly, the TA policy is the solution to 
\begin{align}\label{eq:taopt}
\begin{aligned}
    \min_{\boldsymbol{p}} \ &\sum_{i=1}^N \mathbb{E} [\log h_i \ | R_i = r_i], \\
        \text{s.t.} \ &0 \leq p_i \leq 1, \ \forall \ i,
\end{aligned}
\end{align}
where $R_i$ is a random variable describing the distance of source $i$ from the base station. This is solvable in a distributed fashion because of the separability. Each source independently computes its term in the sum and conditions on its own location, while treating the location of its interferers as random variables. As $N$ goes to infinity, this policy is shown to converge to the PF policy, but simulation results show that even for small networks, the performance is quite similar on average.

The solution to~\eqref{eq:taopt} is given by 
\begin{equation}\label{eq:tasol}
    p_i^{TA} = \Big[ (N-1) \Big( 1 - \frac{1}{r_i^{-2} \theta} \log \big( 1 + r_i^{-2} \theta \big) \Big) \Big]^{-1},
\end{equation}
where the roll-off parameter $\beta$ is assumed to be $2$. This allows for tractable analysis because the expected number of transmitters increases quadratically with $r_i$, while the expected received power decreases quadratically, making the total expected interference from any ring of $\epsilon$ width the same. While this assumption does not generally hold, it provides a good analytical approximation to the true optimal value.

\section{Protocol Design}\label{sec:protocol-design}

The protocols we design and implement in this work are based on IEEE 802.11, but make several important changes to the distributed coordination function (DCF) mechanism. In standard 802.11, the DCF operates using a backoff counter and a contention window. Each source $i$ initializes its contention window $CW_i$ to be $16$, and when it has a packet to send, samples a value in the range $[0,CW_i]$ uniformly at random. The backoff counter is set to this value. The transmitter then senses the channel, and when the channel is idle, it decrements the value of the backoff counter in each slot. If the channel becomes busy at any point, the counter is paused and restarted when the channel becomes idle again. When the counter reaches $0$ and the channel remains idle, the node initiates a transmission.

If the transmission is successful, the size of the contention window is reset to its initial value of $16$ (if it was previously changed). If the transmission is unsuccessful due to a packet collision, the node attempts to retransmit a fixed number of times $\eta$, then doubles the size of its contention window. This mechanism is designed so that a congested channel will not continue to see collisions, and instead each node will ``back off'' its transmission attempts. The near-universal adoption of the 802.11 protocol speaks to the utility of this mechanism. It falls short, however, when dealing with spatial diversity in the network, and suffers from the same unfairness highlighted in~\cite{celik2009mac} and~\cite{jones2023minimizing}. In fact, the backoff mechanism creates a positive feedback loop, amplifying this unfairness even further. Consider two nodes which transmit simultaneously, where one is much farther from the base station than the other. The closer transmission's signal will be much stronger at the base station and will drown out the farther transmission. The farther node will see a collision and back off, amplifying the unfairness in future slots. 

Our protocol addresses this issue by changing the DCF mechanism in two ways. First, it optimizes the size of each node's contention window to mitigate the spatial unfairness which the DCF otherwise causes. Second, it does not resize the contention window upon collisions, because the protocol is actually \textit{designed} so that farther nodes experience more collisions, in order to improve their chances of success and reduce the number of idle slots.

Recall that in the theoretical policies in~\cite{jones2023minimizing}, each source $i$ transmits following a Bernoulli process with a constant probability $p_i^{\pi}$ that is optimized for that particular source under policy $\pi$. By sizing the contention windows appropriately in our protocol, we can approximate this Bernoulli process~\cite{bianchi_performance_2000}. Let the size of the contention window of node $i$ be 
\begin{equation}\label{eq:cwsize}
    CW_i^{\pi} = \frac{2}{p_i^{\pi}} - 2,
\end{equation}
where each time the backoff counter $B_i^{\pi}$ is sampled, it samples uniformly from the range $[0,CW_i^{\pi}]$. The expected backoff counter value is then 
\begin{equation}
    \mathbb{E}[B_i^{\pi}] = \frac{CW_i^{\pi}}{2} = \frac{1}{p_i^{\pi}} - 1.
\end{equation}

Because the counter must reach $0$ before transmitting, the average inter-transmission time is $\mathbb{E}[B_i^{\pi}]+1 = \frac{1}{p_i^{\pi}}$, which is equal to the average inter-transmission time of a Bernoulli process with probability $p_i^{\pi}$. Therefore, the backoff mechanism with a contention window sized according to~\eqref{eq:cwsize} is a good approximation for the theoretical policy $\pi$. In practice, $CW_i^{\pi}$ is often non-integer, so with a slight loss of optimality we simply round to the nearest integer value.

The theoretical policies assume a Rayleigh fading channel and roll-off parameter $\beta$. Our experimental testbed, described in the next section, is indoors and subject to multi-path and other channel effects. Rather than relying on a channel model in our experiments, we directly measure the average received power of each node at the base station, and allow transmitters access to these values. Denote the average received power of node $i$ at the base station as $P_i^{RX}$.

Our protocols then operate in the following way. Under the proportionally fair protocol, called WiFair PF, each source $i$ independently computes its contention window by first solving 
\begin{equation}\label{eq:pfwifairsol}
    \frac{1}{\tilde{p_i}^{PF}} - \sum_{j \in I_i} \frac{1}{1+P_j^{RX}/(P_i^{RX} \theta) - \tilde{p_i}^{PF}} = 0,
\end{equation}
where $r_i^{-\beta}$ in~\eqref{eq:pfopt} has been replaced with the measured average power, and $\theta$ is given by the MCS scheme used in the experiment. Each source then uses~\eqref{eq:cwsize} to find $CW_i^{PF}$, where $p_i^{PF} = \min \{ \tilde{p_i}^{PF}, 1 \}$. It then fixes this contention window size and otherwise operates like standard 802.11, after setting additional parameter values that are specific to each experiment.

The topology agnostic protocol, called WiFair TA, operates in a similar manner, but requires slightly more care to map~\eqref{eq:tasol} from theoretical average power to measured power. Recall that $r_i$ in the theoretical proportionally fair policy is normalized to take values from $0$ to $1$ and assumes $1$ is the maximum distance from the base station where any source can exist. In translating to received power, the analog of this quantity is the smallest average power which any source can have at the receiver. Denote this quantity as $P_{min}^{RX}$. Then $r_i^{-2}$, the average normalized received power in the theoretical model, is equivalent to $\gamma_i = P_i^{RX} / P_{min}^{RX}$.

Following this method, each source $i$ in WiFair TA computes its contention window by plugging~\eqref{eq:tasol} into~\eqref{eq:cwsize}, and replacing $r_i^{-2}$ with $\gamma_i$. This yields 
\begin{equation}
    CW_i^{TA} = 2 (N-1) \Big( 1 - \frac{1}{\gamma_i \theta} \log \big( 1 + \gamma_i \theta \big) \Big) - 2.
\end{equation}
Just as in WiFair PF, the protocol fixes its contention windows to these values, sets any additional parameter values, and otherwise operates like standard 802.11. In the next section, we describe the experimental setup used to test these protocols.


\section{Experimental Setup}

\begin{figure}
    \centering
    \includegraphics[width=0.5\textwidth]{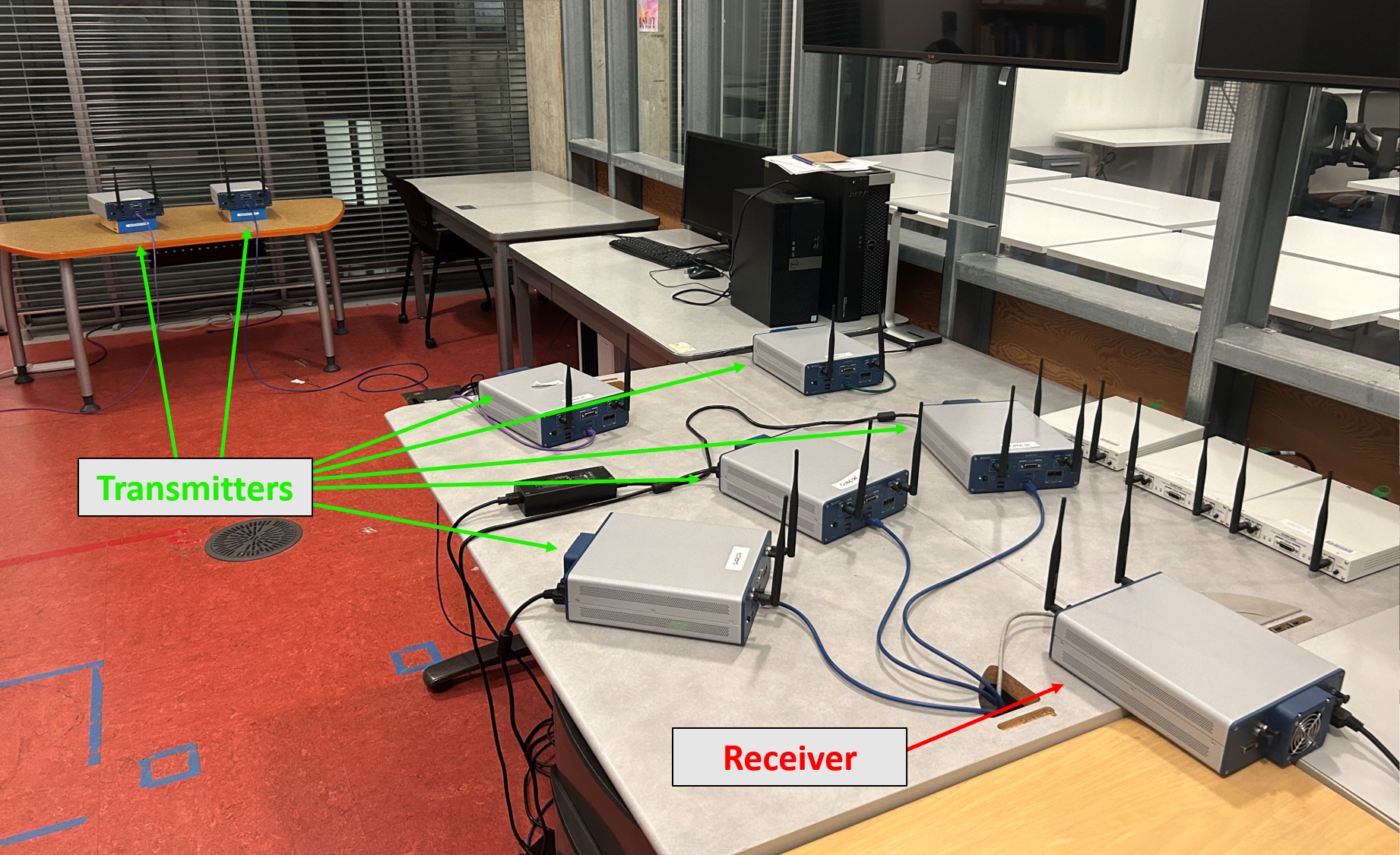}
    \caption{Experimental setup with seven transmitters and one receiver}
    \label{fig:lab-setup}
\end{figure}

We implemented our protocols on a testbed of eight NI USRP-2974 SDRs, with seven acting as transmitters and one acting as the base station. The transmitters were placed in three rings at varying distances from the base station to create spatial diversity in the network. Each radio was programmed with LabVIEW NXG software, using a modified version of the 802.11 implementation included in the LabVIEW Communication Systems Design Suite. We also made extensive use of the modifications for AoI measurements and calculations implemented in~\cite{kadota2021wifresh}.

As mentioned above, the average received power from each transmitter was measured before each experiment, and the transmission power of each source was then adjusted to match the desired power level. These levels are included in Table~\ref{table:tx_placement} in units of decibels relative to full scale (dBFS), which represent the power relative to the maximum allowable level in the digital hardware before clipping.

\begin{table}[!h]
\begin{center}
\begin{tabular}{ | c | >{\centering\arraybackslash}m{2cm} | >{\centering\arraybackslash}m{2cm} | >{\centering\arraybackslash}m{1cm} |} 
  \hline
  Transmitter & Avg Rx Power (dBFS) \\
  \hline
  1,2,3 & -15 \\ 
  \hline
  4,5 & -33 \\ 
  \hline
  6,7 & -40 \\ 
  \hline
\end{tabular}
\end{center}
\caption{Transmitter Placement}
\label{table:tx_placement}
\end{table}

The AoI of each node was measured using the LabVIEW code developed in~\cite{kadota2021wifresh}, which appends a timestamp of the current time to each packet when it is generated. The base station keeps track of the timestamp of the last received packet from each transmitter, and computes the AoI as the difference between the current time and this timestamp.

The MAC layer, including this functionality, was implemented on an FPGA in each USRP with a $10$ MHz clock, and thus operates on the order of microseconds. To account for synchronization issues and possible drift, each USRP periodically synchronized its operating system time using the Network Time Protocol (NTP), and used this time to compute timestamps. As a secondary precaution, after the AoI of each source was sampled in an experiment, the values of each source were normalized so that the minimum AoI value across the entire experiment was zero. The propagation and processing delay was again on the order of microseconds, so the error from discounting these effects was orders of magnitude smaller than the measured AoI, as we will see in the next section. Moreover, because all radios were normalized in the same way, any possible error affected all transmitters equally.

The DCF is handled entirely in the MAC layer on the FPGA, so implementing the fixed contention windows in our protocols required reprogramming the FPGA of each radio. Within the FPGA, the backoff timer is sampled using a shift register that generates pseudo-random numbers between $0$ and $1023$, the maximum contention window size in 802.11. Generating pseudo-random numbers between $0$ and $2^n - 1 \leq 1023$ for some integer $n$ can be accomplished by applying a bit mask to the original value, and because 802.11 only uses contention windows that are powers of $2$, this is how it is done in practice.

Our protocol requires carefully sized contention windows which are not necessarily powers of $2$. To generate a random backoff timer in this interval, we first AND the original pseudo-random number with the smallest value $2^n - 1$ which is at least as large as our contention window $CW$. If the result is no larger than $CW$, we accept it. Otherwise, we take the previous backoff value and increment it by $1$ with probability $0.5$, else we decrement it by $1$. If this value is between $0$ and $CW$, we accept it, else we use the previous backoff value. While not truly uniform or independent across time, this approach yields an approximately uniform distribution, which can be sampled in a few clock cycles of the FPGA.

Before each experiment, the average power levels of each source were measured, and their transmission powers were adjusted so that the average power at the receiver matched Table~\ref{table:tx_placement}. With these power levels, the optimal contention windows under WiFair PF and WiFair TA were computed using the method in Section~\ref{sec:protocol-design} and are listed in Table~\ref{table:cw-sizes}. These values were used in all three experiments.

\begin{table}[!h]
    \begin{center}
    \begin{tabular}{ | c | >{\centering\arraybackslash}m{1.5cm} | >{\centering\arraybackslash}m{1.5cm} | >{\centering\arraybackslash}m{1.5cm} |} 
      \hline
      \textbf{Source} & \textbf{WiFair PF} & \textbf{WiFair TA} \\
      \hline
      1,2,3 & 11 & 10 \\
      \hline
      4,5 & 5 & 9 \\
      \hline
      6,7 & 3 & 7 \\
      \hline
    \end{tabular}
    \end{center}
    \caption{Optimal contention windows for each source}
    \label{table:cw-sizes}
\end{table}

The minimum receive power $P_{min}^{RX}$, used to compute the WiFair TA contention windows, was set to $-45$ dB. This was the lowest receive power sufficiently above the noise floor that the receiver was able to decode packets at a high success rate without interference.

Each experiment was run in the ISM frequency band at 2.4 GHz, using a 20 MHz channel. The 802.11 RTS/CTS mechanism was disabled for all experiments, which is standard for small packet sizes in most 802.11 implementations. One challenge in using our small testbed was that 802.11 only begins to see AoI performance degrade when the network is congested and collisions occur. With only $7$ transmitters, we needed to create additional congestion to truly see these effects. We accomplished this by decreasing the default contention window size from $16$ to $8$ when testing 802.11. Additional parameters were experiment-specific and are detailed in the next section.



\section{Experimental Results}

\subsection{Single-Packet Queues}

In our first experiment, we followed the theoretical model in~\cite{jones2023minimizing} as closely as possible by using LCFS single-packet queues. In this setting, each packet is assumed to represent a sensor update, and the base station is assumed to only care about the most recent update. Therefore, all but the most recent packet are discarded before transmitting. Each transmitter generated packets of random data with a length of $100$ bytes and a generation rate of $10,000$ packets per second. This rate was set large enough so that a packet was always available to send, but because of the queueing mechanism, there was never a backlog. 

The MCS was set to $0$, which corresponds to a $1/2$ rate coding scheme and BPSK modulation. This low-data-rate MCS is again appropriate for the setting of small update packets. From~\cite{snr2024}, the required SIR threshold $\theta$ for this MCS is $5 dB \approx 3.16$. In addition, the number of transmission retries after a collision $\eta$ was set to $0$. This ensured that each source sent the most recent update after a collision instead of attempting to re-send a stale packet.

\begin{figure}
    \centering
    \includegraphics[width=0.5\textwidth]{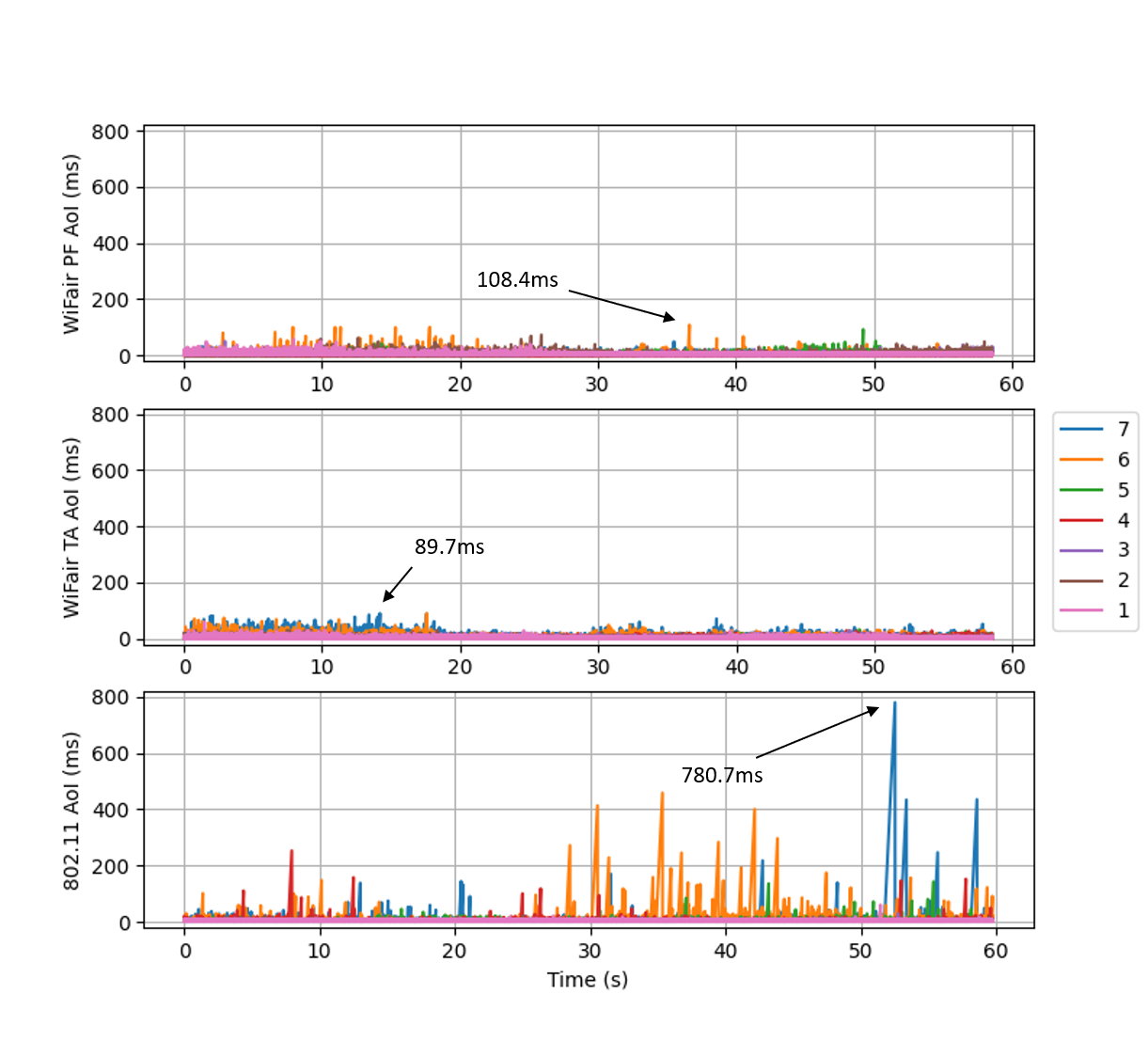}
    \caption{AoI under WiFair vs 802.11 with single-packet LCFS queues}
    \label{fig:aoi-lcfs}
\end{figure}

We tested both WiFair PF and WiFair TA against 802.11 (also run with LCFS single-packet queues), using this setup and the contention windows listed in Table~\ref{table:cw-sizes}. The system was run for approximately $60$ seconds, and a plot of the AoI of each source over time under each of the three protocols is shown in Figure~\ref{fig:aoi-lcfs}, with the peak values labeled in the plot. The mean AoI of each source, and the network average AoI, is listed in Table~\ref{table:aoi-lcfs}.

\begin{table}[!h]
    \begin{center}
    \begin{tabular}{ | c | >{\centering\arraybackslash}m{1.5cm} | >{\centering\arraybackslash}m{1.5cm} | >{\centering\arraybackslash}m{1.5cm} |} 
      \hline
      \textbf{Source} & \textbf{WiFair PF} & \textbf{WiFair TA} & \textbf{802.11} \\
      \hline
      1 & 5.18 & 4.70 & 3.85 \\
      \hline
      2 & 6.57 & 4.36 & 3.93 \\
      \hline
      3 & 5.05 & 3.94 & 3.55 \\
      \hline
      4 & 3.90 & 4.33 & 5.79 \\
      \hline
      5 & 5.24 & 3.79 & 5.05 \\
      \hline
      6 & 5.78 & 6.74 & 16.52 \\
      \hline
      7 & 4.07 & 9.05 & 14.95 \\
      \hline
      \textbf{Avg} & \textbf{5.11} & \textbf{5.27} & \textbf{7.66} \\
      \hline 
    \end{tabular}
    \end{center}
    \caption{Mean AoI in ms for LCFS single-packet queues}
    \label{table:aoi-lcfs}
\end{table}

The results show that both WiFair protocols behave roughly the same on average, with an average AoI of $33\%$ and $32\%$ below 802.11 respectively. The more notable improvements, however, are the reduction in peak AoI and the fairness seen across each source. The peak AoI values seen under WiFair were $\textit{86\%}$ and $\textit{89\%}$ lower than 802.11 repectively. While average AoI is an important metric, large spikes in AoI correspond to times where a system has very outdated information, and can make very incorrect decisions. Therefore, the improvement in peak AoI is notable.

The average AoI of each source under 802.11 increases as the source gets farther from the base station. This supports the spatial unfairness that we expect to see, with the farthest sources seeing an AoI $4$ to $5$ times as large as the closest sources. In addition, a quick examination of Figure~\ref{fig:aoi-lcfs} shows that nearly every notable peak belongs to sources $6$ and $7$.

Alternatively, under WiFair PF, there is no discernible spatial unfairness. In fact, source $7$ has the second lowest AoI of any source. WiFair TA sees slightly worse results in terms of fairness. This is due to the assumption that nodes are uniformly distributed in space, so nodes $6$ and $7$ assume that there are nodes even farther away, whereas in WiFair PF they know that they are the farthest. Source $7$ also performs slightly worse than source $6$ in WiFair TA, likely due to channel effects at the time of the experiment.

Overall, these results verify that spatial unfairness exists in 802.11 and that WiFair mitigates this unfairness, dramatically reducing peak AoI, and bringing down the network average AoI. This supports the theoretical claims in~\cite{jones2023minimizing} and the usefulness of WiFair as a practical protocol in the LCFS single-packet queue setting. In the next experiments, we show that even in different settings which deviate from the theoretical model, WiFair still performs remarkably well.

\subsection{FCFS Queues with Video Streams}

Because the theoretical policies are designed for the LCFS single-packet queue scenario in the previous section, it is reassuring but not entirely surprising that their performance exceeds 802.11. To test the robustness of our protocol, we also experimented with traditional FCFS queues.

We again set the MCS to $0$, corresponding to a $1/2$ rate code and BPSK modulation, and an SIR threshold value $\theta \approx 3.16$. The retransmission attempt paramter $\eta$ was set to $4$, which is the default value in Labview's 802.11 implementation, to prevent packet loss in the event of a collision.

We generated our FCFS data using a video file at each source, which we streamed to Labview over a UDP connection using the ffmpeg library with a packet size of $1500$ bytes. The stream simulated a live video stream, as it was sent using a real-time connection. Labview then read the data from the stream in real-time and sent it over the air using the protocol currently active. At the receiver, in a similar fashion, Labview received the data over the air, identified which source each packet originated from, and sent it over another live UDP stream corresponding to the appropriate source. This stream was received by ffmpeg and written to a file as it was received. Then, in addition to monitoring the AoI, we were able to watch the received video file from each source. 

The addition of real data into the experiment introduced another variable of data rate. As data rate increases, the network becomes more congested, and we expect all three protocols to perform worse. While sending the UDP video stream without additional error correction, we observed a critical point in the data rates, where once a certain level of congestion was reached, the video quality would degrade sharply. For this experiment, we operated with relatively low data rates, below this critical point. The video file sent from each source was $480$p and had a bitrate of $270$ kbps, resulting in a total data rate of $1890$ kbps from all seven sources.

\begin{figure}[htbp]
    \centering
    \includegraphics[width=0.5\textwidth]{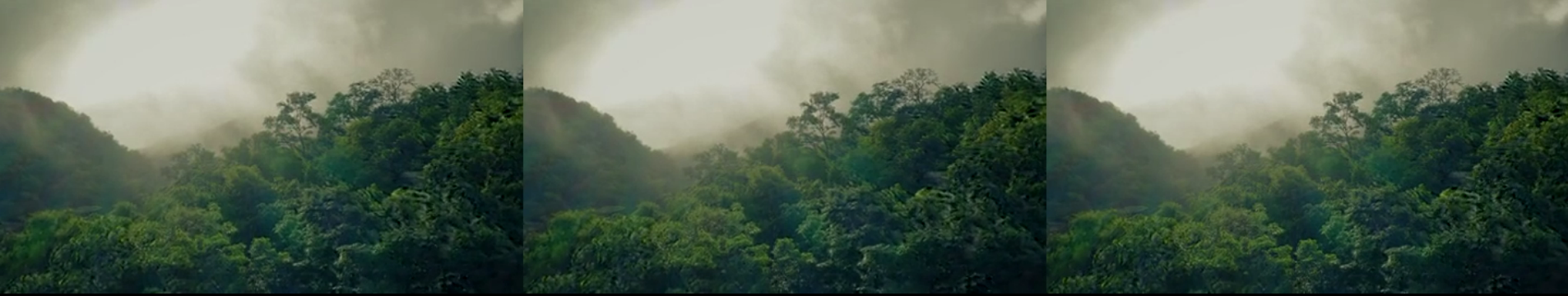}
    \caption{Received video screenshot under WiFair PF, WiFair TA, and 802.11 respectively \protect\footnotemark}
    \label{fig:video-screenshot}
\end{figure}

The received video quality was roughly the same between all three protocols. Due to the lack of error correction, there were occasional artifacts that appeared in the videos from almost every source and under each protocol, but the videos were mostly clear. Figure~\ref{fig:video-screenshot} shows a screenshot of three videos, which originated from the same source and were sent using each of the three protocols.

\begin{figure}
    \centering
    \includegraphics[width=0.5\textwidth]{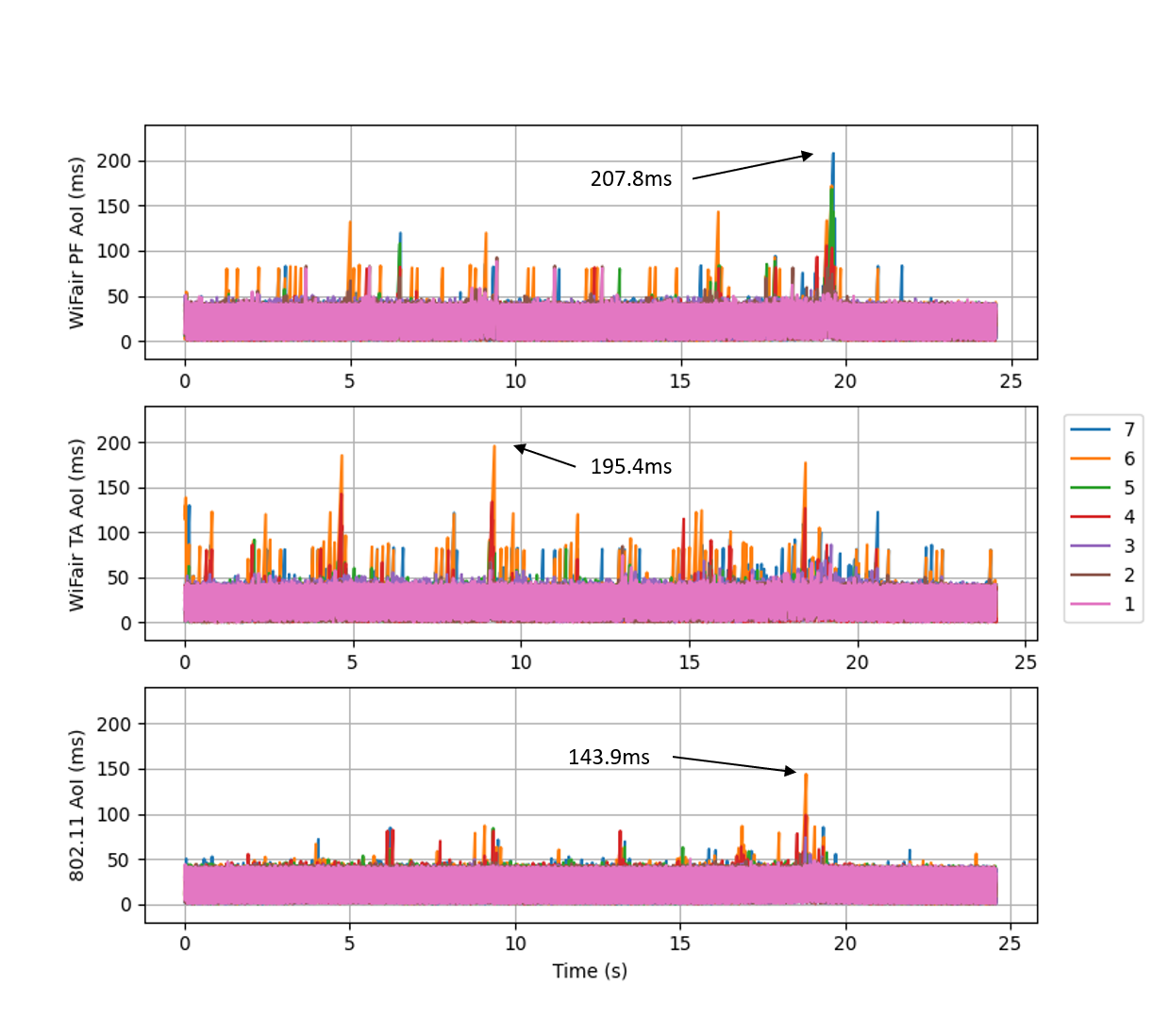}
    \caption{AoI under WiFair vs 802.11 with FCFS queues sending UDP video streams}
    \label{fig:aoi-fcfs-udp}
\end{figure}

The plot of AoI over time with the peak values listed is shown in Figure~\ref{fig:aoi-fcfs-udp}. Likewise, the average AoI of each source is shown in Table~\ref{table:aoi-fcfs-udp}. At this level of congestion, there is no notable difference in performance between the three protocols. 802.11, in fact, showed the lowest average AoI and the lowest peak AoI. Intuitively, this makes sense. When congestion is low, collisions are rare, and secondary collisions are extremely rare. This means that contention windows will not grow too large, and the spatial unfairness seen in the first experiment is no longer present. WiFair, by imposing fixed contention windows, allows for a higher chance of secondary collisions, which explains the slightly worse performance.

\footnotetext{Stock footage provided by Videvo, downloaded from www.videvo.net}

\begin{table}[!h]
    \begin{center}
    \begin{tabular}{ | c | >{\centering\arraybackslash}m{1.5cm} | >{\centering\arraybackslash}m{1.5cm} | >{\centering\arraybackslash}m{1.5cm} |} 
      \hline
      \textbf{Source} & \textbf{WiFair PF} & \textbf{WiFair TA} & \textbf{802.11} \\
      \hline
      1 & 22.14 & 22.74 & 20.91 \\
      \hline
      2 & 23.30 & 21.91 & 21.42 \\
      \hline
      3 & 22.92 & 24.66 & 21.19 \\
      \hline
      4 & 22.13 & 23.48 & 22.51 \\
      \hline
      5 & 23.02 & 24.18 & 22.33 \\
      \hline
      6 & 25.40 & 30.08 & 23.42 \\
      \hline
      7 & 23.99 & 28.79 & 22.37 \\
      \hline
      \textbf{Avg} & \textbf{23.27} & \textbf{25.12} & \textbf{22.02} \\
      \hline
    \end{tabular}
    \end{center}
    \caption{Mean AoI in ms for FCFS queues sending UDP video streams}
    \label{table:aoi-fcfs-udp}
\end{table}

\subsection{FCFS Queues with Congestion}

In our final experiment, we tested the performance of WiFair using FCFS queues under more significant congestion. We set the MCS to $5$, corresponding to a $2/3$ rate code and a $64$-QAM constellation. From~\cite{snr2024}, this requires an SIR threshold $\theta = 17.5 dB \approx 56.23$. We generated $200$ byte packets of random data, with each source generating $600$ packets per second. This resulted in a data rate of $960$ kbps from each source, and a total data rate of $6.72$ Mbps in the network. This additional traffic created the congestion needed to test WiFair with FCFS queues in a more interesting setting than our second experiment. The retransmission attempt parameter $\eta$ was again set to $4$.

We ran this experiment slightly differently, allowing each source to run for a bit of time before collecting data. We did this because each queue is initially empty, and we wanted the queues to reach a steady state before measuring AoI. After this initial ramp-up, we collected data for roughly $120$ seconds under both WiFair TA and 802.11.

\begin{figure}
    \centering
    \includegraphics[width=0.5\textwidth]{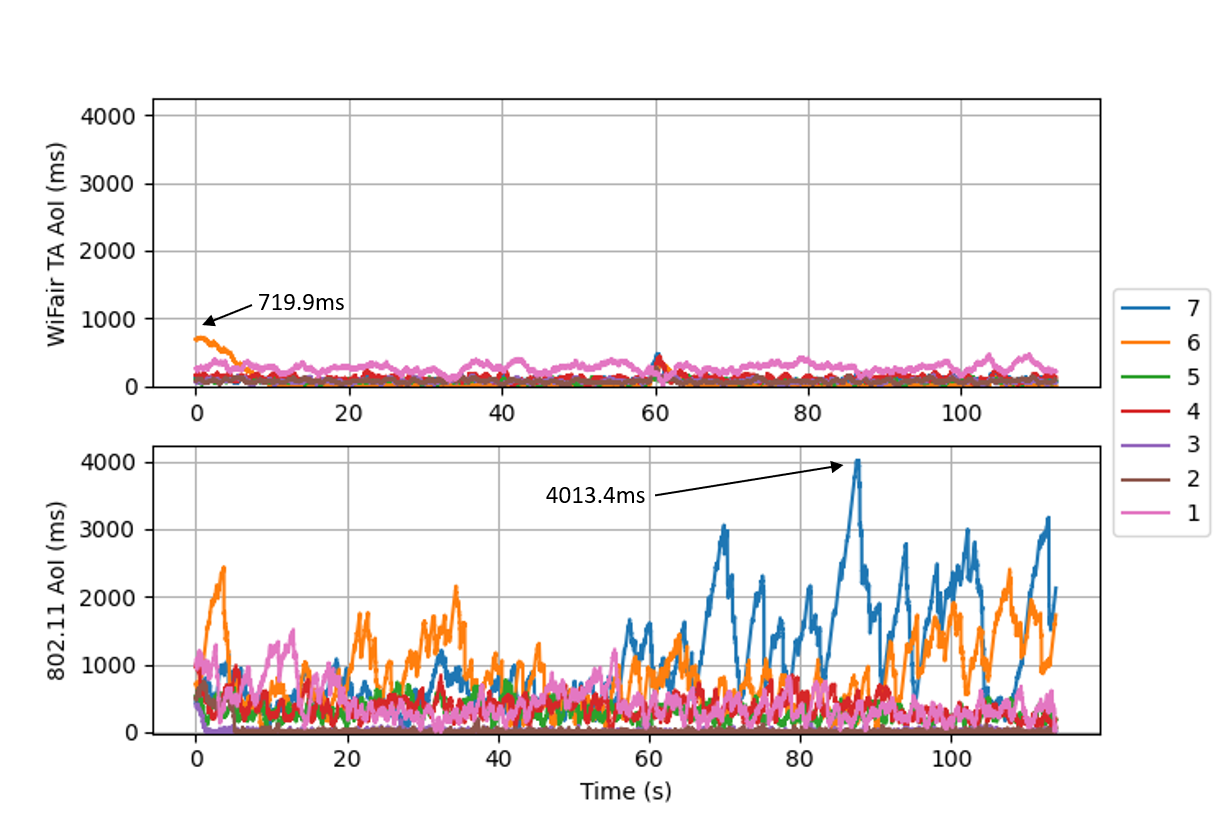}
    \caption{AoI under WiFair TA vs 802.11 with FCFS queues and congestion}
    \label{fig:aoi-fcfs}
\end{figure}

Figure~\ref{fig:aoi-fcfs} shows the plot of AoI over time for each source, with the peak value labeled, and Table~\ref{table:aoi-fcfs} shows the average AoI of each source. The performance difference here is striking. Not only is the network average AoI $76\%$ lower under WiFair, the peak AoI is $\textit{82\%}$ lower.

Examining individual sources, the performance difference becomes even more stark. The introduction of queueing and congestion causes WiFair to lose some of its fairness properties, and the average AoI of each source is more varied than in the single-packet case or the FCFS case with limited congestion. However, the differences between nodes are not correlated with their location, and node $6$ in fact has the lowest average AoI. 802.11 shows a far more drastic disparity between nodes, with a factor of over $\textit{80}$ times difference in the AoI of nodes $3$ and $7$. Furthermore, as in the previous experiments, the spatial unfairness causes nodes $6$ and $7$ to have by far the largest average AoI, and the most notable peaks in AoI in Figure~\ref{fig:aoi-fcfs}. The AoI of node $7$ in this experiment peaked at \textit{over $4$ seconds}.

The results here clearly show that, while queueing introduces an additional level of complexity, and WiFair cannot promise the same theoretical results as in the single-packet queue case, it is robust enough to still perform well. It still solves the spatial unfairness present in 802.11, and it yields significantly lower average and peak AoI.

\begin{table}[!h]
    \begin{center}
    \begin{tabular}{ | c | >{\centering\arraybackslash}m{1.5cm} | >{\centering\arraybackslash}m{1.5cm} | >{\centering\arraybackslash}m{1.5cm} |} 
      \hline
      \textbf{Source} & \textbf{WiFair TA} & \textbf{802.11} \\
      \hline
      1 & 265.97 & 428.03 \\
      \hline
      2 & 71.13 & 33.05 \\
      \hline
      3 & 71.24 & 13.02 \\
      \hline
      4 & 117.92 & 368.44 \\
      \hline
      5 & 74.45 & 303.21 \\
      \hline
      6 & 51.11 & 836.91 \\
      \hline
      7 & 85.72 & 1059.37 \\
      \hline
      \textbf{Avg} & \textbf{105.36} & \textbf{434.58} \\
      \hline
    \end{tabular}
    \end{center}
    \caption{Mean AoI in ms with FCFS queues and congestion}
    \label{table:aoi-fcfs}
\end{table}

In Table~\ref{table:experiment-params}, we summarize the parameters used in the experiments. All three experiments motivate the use of WiFair as a practical protocol for minimizing AoI and eliminating the spatial unfairness present in 802.11. It keeps average and especially peak AoI small, and it is robust enough to handle a variety of different traffic types and queueing methods. It shows no significant signs of spatial unfairness, and it scales much more gracefully with the level of congestion in the network compared to 802.11. 

\begin{table}[!h]
    \begin{center}
    \begin{tabular}{ | c | >{\centering\arraybackslash}m{1.5cm} | >{\centering\arraybackslash}m{1.5cm} | >{\centering\arraybackslash}m{1.5cm} |} 
      \hline
      \textbf{Parameter} & \textbf{Exp 1} & \textbf{Exp 2} & \textbf{Exp 3} \\
      \hline
      Queueing discipline & LCFS SP & FCFS & FCFS \\
      \hline
      Data source & Random & UDP stream & Random \\
      \hline
      Packet size (bytes) & 100 & 1500 & 200 \\
      \hline
      Data rate (kbps) & At will & 1890 & 6720 \\
      \hline
      MCS & 0 & 0 & 5 \\
      \hline
      Retransmission atttempts & 0 & 4 & 4 \\
      \hline
    \end{tabular}
    \end{center}
    \caption{Summary of experiment parameters}
    \label{table:experiment-params}
\end{table}

\section{Conclusion}

In this work, we design and implement two random access protocols on an SDR testbed, designed to minimize AoI and mitigate spatial unfairness that exists in standard 802.11 and other protocols. We draw on our recent theoretical results along with the practical aspects of 802.11 to design effective and practical protocols. We test their performance against 802.11 and show that they achieve significant improvements in average AoI, nearly an order of magnitude improvement in peak AoI, and eliminate nearly all signs of spatial unfairness in both the LCFS single-packet queue case and the FCFS queue case with congestion. Moreover, we show that WiFair is able to serve more bursty traffic by demonstrating a live video stream, further proving its robustness. Future work involves modifying the protocol to allow for dynamically resizing networks and mobile nodes.

\bibliographystyle{IEEEtran}
\bibliography{wifair}

\end{document}